\documentclass[8pt,twocolumn]{article}

\usepackage[pdftex]{graphicx}
\usepackage{amssymb,amsfonts,amsmath}
\usepackage{cite}
\usepackage[auth-sc]{authblk}
\usepackage[top=1in, bottom=1in, left=1in, right=1in]{geometry}
\usepackage{hyperref}

\newlength{\HalfPage}
\begin{document}

\title{Predator confusion is sufficient to evolve swarming behavior}

\author[1,4]{\small Randal S.~Olson}
\author[2,4]{Arend Hintze}
\author[3,4]{Fred C.~Dyer}
\author[2,4]{David B.~Knoester}
\author[2,4]{Christoph Adami}

\affil[1]{Department of Computer Science and Engineering, Michigan State University, East Lansing, MI 48824, USA}
\affil[2]{Department of Microbiology and Molecular Genetics, Michigan State University, East Lansing, MI 48824, USA}
\affil[3]{Department of Zoology, Michigan State University, East Lansing, MI 48824, USA}
\affil[4]{BEACON Center for the Study of Evolution in Action, Michigan State University, East Lansing, MI 48824, USA}

\date{}


\maketitle

Swarming behaviors in animals have been extensively studied due to their implications for the evolution of cooperation, social cognition, and predator-prey dynamics. An important goal of these studies is discerning which evolutionary pressures favor the formation of swarms. One hypothesis is that swarms arise because the presence of multiple moving prey in swarms causes confusion for attacking predators, but it remains unclear how important this selective force is. Using an evolutionary model of a predator-prey system, we show that predator confusion provides a sufficient selection pressure to evolve swarming behavior in prey. Furthermore, we demonstrate that the evolutionary effect of predator confusion on prey could in turn exert pressure on the structure of the predator's visual field, favoring the frontally oriented, high-resolution visual systems commonly observed in predators that feed on swarming animals. Finally, we provide evidence that when prey evolve swarming in response to predator confusion, there is a change in the shape of the functional response curve describing the predator's consumption rate as prey density increases. Thus, we show that a relatively simple perceptual constraint---predator confusion---could have pervasive evolutionary effects on prey behavior, predator sensory mechanisms, and the ecological interactions between predators and prey.

\hspace{0.1in}

{\parindent0pt Keywords: {\it swarming behavior}, {\it predator confusion effect}, {\it predator-prey coevolution}, {\it predator visual system}, {\it functional response}}

\section{Introduction}

The sudden emergence of a cohesive swarm from the behavioral decisions of individual animals is one of nature's most striking examples of collective animal behavior~\cite{Couzin2009}. For example, European starlings ({\it Sturnus vulgaris}) are known to form spectacular, coordinated flocks composed of hundreds of thousands of birds, seemingly without any form of leadership~\cite{Feare1984,Hemelrijk2011}. During their monthly breeding seasons, Atlantic herring ({\it Clupea harengus}) aggregate into schools comprising hundreds of millions of fish to spawn offspring~\cite{Makris2009}. Perhaps most notoriously, desert locusts ({\it Schistocerca gregaria}) form massive swarms with billions of locusts that devastate entire agricultural zones in Africa, the Near East, and southwest Asia~\cite{Symmons2001}.

While swarm-like aggregations could arise for relatively simple reasons, e.g., to converge on a common resource~\cite{Kersten1991}, in many cases swarms are formed via behavioral mechanisms that coordinate the movements of individuals to ensure group cohesion~\cite{Ballerini2008}. Since swarming may incur a variety of fitness costs (e.g., increased attack rate from predators on larger swarms), considerable effort has been devoted to understanding the compensatory benefits of swarming~\cite{Krause2002}. Many such benefits have been proposed: Swarming may improve mating success~\cite{Diabate2011,Yuval1993}, increase foraging efficiency~\cite{Pulliam1984}, and provide distributed information processing abilities~\cite{Couzin2009}. In this study, we focus on swarming as a defense against predation~\cite{Krause2002}.

Evolved swarm behaviors could protect group members from predators in several ways. For example, swarming improves group vigilance~\cite{Treherne1981,Kenward1978,Pulliam1973,Treisman1975}, reduces the chance of being encountered by predators~\cite{Treisman1975,Inman1987}, dilutes an individual's risk of being attacked~\cite{Treherne1982,Foster1981,Hamilton1971}, enables an active defense against predators~\cite{Bertram1978}, and reduces predator attack efficiency by confusing the predator~\cite{Ioannou2008,Jeschke2007}. Given the long generation times of many of the animals involved (months to years), it is exceedingly difficult to discern which of these benefits, if any, are sufficient to produce swarming as an evolutionary response, let alone study the properties of swarm behaviors as they evolve~\cite{Jeschke2007,Beauchamp2004}.


To address this challenge, we investigate the evolutionary origins of swarming behavior in a digital system. Digital systems have previously been used to provide key insights into core evolutionary processes~\cite{Lenski2003,Wilke2001}, and several well-known studies have adopted digital systems as a method to study swarm behavior~\cite{Lukeman2010,Couzin2005,Couzin2002}. More recently, digital systems have even been used to elucidate the emergence of prey swarming behavior as a response to predation~\cite{Ioannou2012}. These previous studies have provided insight into the fundamental dynamics of swarming behavior. However, most have not focused on isolating the evolutionary pressures that might favor the formation of swarms, and none have explored the coevolution of predator and prey behavior. In fact, except for only a handful of studies~\cite{Olson2013,Tosh2011,Wood2007,Reluga2005,Ward2001}, this literature typically has not studied Darwinian evolution as a process affecting the properties of swarms. Here, we present a model in which predators and groups of genetically homogeneous prey are coevolved in a two-dimensional virtual environment. Predators are endowed with a retina that enables them to observe prey, while prey are equipped with a retina that enables them to sense both conspecifics and predators. In this model, predator and prey are preferentially selected based on how effective they are at consuming prey and surviving, respectively. Swarming is a possible solution for the prey, but is not selected for directly.

While there are many different selective pressures that have been hypothesized to produce swarming behavior, within this digital environment we specifically study the evolution of swarming in the presence of \emph{predator confusion}. In the predator confusion hypothesis, the presence of multiple individuals moving in a swarm confuses approaching predators, making it difficult for them to successfully execute an attack~\cite{Ioannou2008,Jeschke2007,Krakauer1995,Humphries1970}. In a recent review of predator-prey systems with swarming prey, Jeschke and Tollrian noted that predators appeared to become confused by swarming behavior in 16 of the 25 systems reviewed~\cite{Jeschke2007}. However, evidence that predator confusion is a seemingly widespread phenomenon still leaves open the question of how effective predator confusion could be as a selective force favoring the evolution of swarming behavior.

Predator confusion is broadly interesting for two additional reasons. First, it provides an opportunity to study how swarming behavior can in turn exert evolutionary pressures on predators, especially on the perceptual constraints that allow for predator confusion in the first place. For example, once swarming behavior evolves in prey, predator confusion may in turn provide a selective advantage for predators that are no longer confused by swarms. Second, predator confusion may influence the \emph{functional response} describing the predator's consumption rate as prey density increases~\cite{Holling1959}, as suggested in a previous study~\cite{Jeschke2005}. Understanding how pervasive mechanisms such as predator confusion affect functional response relationships is critical for accurately modeling the dynamics of predator-prey interactions over ecological and evolutionary time~\cite{Hairston2005}.

The contributions of this work are as follows. First, we demonstrate that the predator confusion effect provides a sufficient selective advantage for prey to evolve swarming behavior. Furthermore, given prey that swarm as a result of the predator confusion effect, we show that predators could in turn be selected to evolve a frontally-oriented, high-resolution visual field. Finally, we provide evidence that the shape of the predator functional response curve can be affected when prey evolve swarming behavior in response to the predator confusion effect. Consequently, we demonstrate that predator confusion could have extensive evolutionary effects on traits ranging from prey behavior to predator sensory mechanisms, as well as the ecological interactions between predators and prey.

\section{Methods}

To study the effects of predator confusion on the evolution of swarming, we create an agent-based simulation in which predator and prey agents interact in a continuous two-dimensional virtual environment. Each agent is controlled by a {\em Markov Network} (MN), which is a stochastic state machine that makes control decisions based on a combination of sensory input (i.e., vision) and internal states (i.e., memory)~\cite{Edlund2011}. We coevolve the MNs of predators and prey with a genetic algorithm, selecting for MNs that exhibit behaviors that are more effective at consuming prey and surviving, respectively. Certain properties of the sensory and motor behavior of predators and prey are implemented as constraints that model some of the differences between predators and prey observed in nature (e.g., relative movement speed, turning agility, and, for predators, maximum consumption rate). Predator confusion, described in more detail below, is implemented as a constraint on predator perception that can be varied experimentally. The source code\footnote{Code: \href{https://github.com/adamilab/eos}{https://github.com/adamilab/eos}} and data\footnote{Data: \href{https://github.com/adamilab/eos-data}{https://github.com/adamilab/eos-data}} for these experiments are available online. In the remainder of this section, we summarize the evolutionary process that enables the coevolution of predator and prey, describe the sensory-motor architecture of individual agents, then present the characteristics of the environment in which predator and prey interact. A detailed description of MNs and how they are evolved can be found in the SI text. 

\subsection{Coevolution of predator and prey}

We coevolve the predator and prey with a \emph{genetic algorithm} (GA), which is a digital model of evolution by natural selection~\cite{Goldberg1989}. In a GA, pools of genomes are evolved over time by evaluating the fitness of each genome at each generation and preferentially selecting those with higher fitness to populate the next generation. The genomes here are variable-length strings of integers that are translated into MNs during fitness evaluation (see SI text).


To perform this coevolution, we create separate genome pools for the predator and prey genomes. Next, we evaluate the genomes' fitness by selecting pairs of predator and prey genomes at random without replacement, then place each pair into a simulation environment and evaluate them for 2,000 simulation time steps.  Within this simulation environment, we generate 50 identical prey agents from the single prey genome and compete them with the single predator agent to obtain their respective fitness. This evaluation period is akin to the agents' lifespan, hence each agent has a potential lifespan of 2,000 time steps (enough time for the prey to travel approximately 400 body lengths). The fitness values, calculated using the fitness function described below, are used to determine the next generation of the respective genome pools. Parameters describing the operation of this GA are summarized in Table S1. At the end of the lifetime simulation, we assign the predator and prey genomes separate fitness values according to the fitness functions:

\begin{equation}
W_{{\rm predator}} = \sum_{t=1}^{2,000} S - A_{t}
\label{eq:pred-fitness}
\end{equation}

\begin{equation}
W_{{\rm prey}} = \sum_{t=1}^{2,000} A_{t}
\label{eq:prey-fitness}
\end{equation}
where $t$ is the current simulation time step, $S$ is the starting swarm size (here, $S$ = 50), and $A_{t}$ is the number of prey agents alive at simulation time step $t$. It can be shown that the predator fitness (Eq.~\ref{eq:pred-fitness}) is proportional to the mean kill rate $k$ (mean number of prey consumed per time step), while the prey fitness (Eq.~\ref{eq:prey-fitness}) is proportional to $(1 - k)$. Thus, predators are awarded higher fitness for capturing more prey faster, and prey are rewarded for surviving longer. We only simulate a portion of the prey's lifespan where they are under predation because we are investigating swarming as a response to predation, rather than a feeding or mating behavior.

Once we evaluate all of the predator-prey genome pairs in a generation, we perform fitness-proportionate selection on the populations via a Moran process, allow the selected genomes to asexually reproduce into the next generation's populations, increment the generation counter, and repeat the evaluation process on the new populations until the final generation (1,200) is reached. 


We perform 180 replicates of each experiment, where for each replicate we seed the prey population with a set of randomly-generated MNs and the predator population with a pre-evolved predator MN that exhibits rudimentary prey-tracking behavior. Seeding the predator population in this manner only serves to speed up the coevolutionary process, and has negligible effects on the outcome of the experiment (Figure S1). 

\begin{figure}[tb]
\centering
\includegraphics[width=\HalfPage]{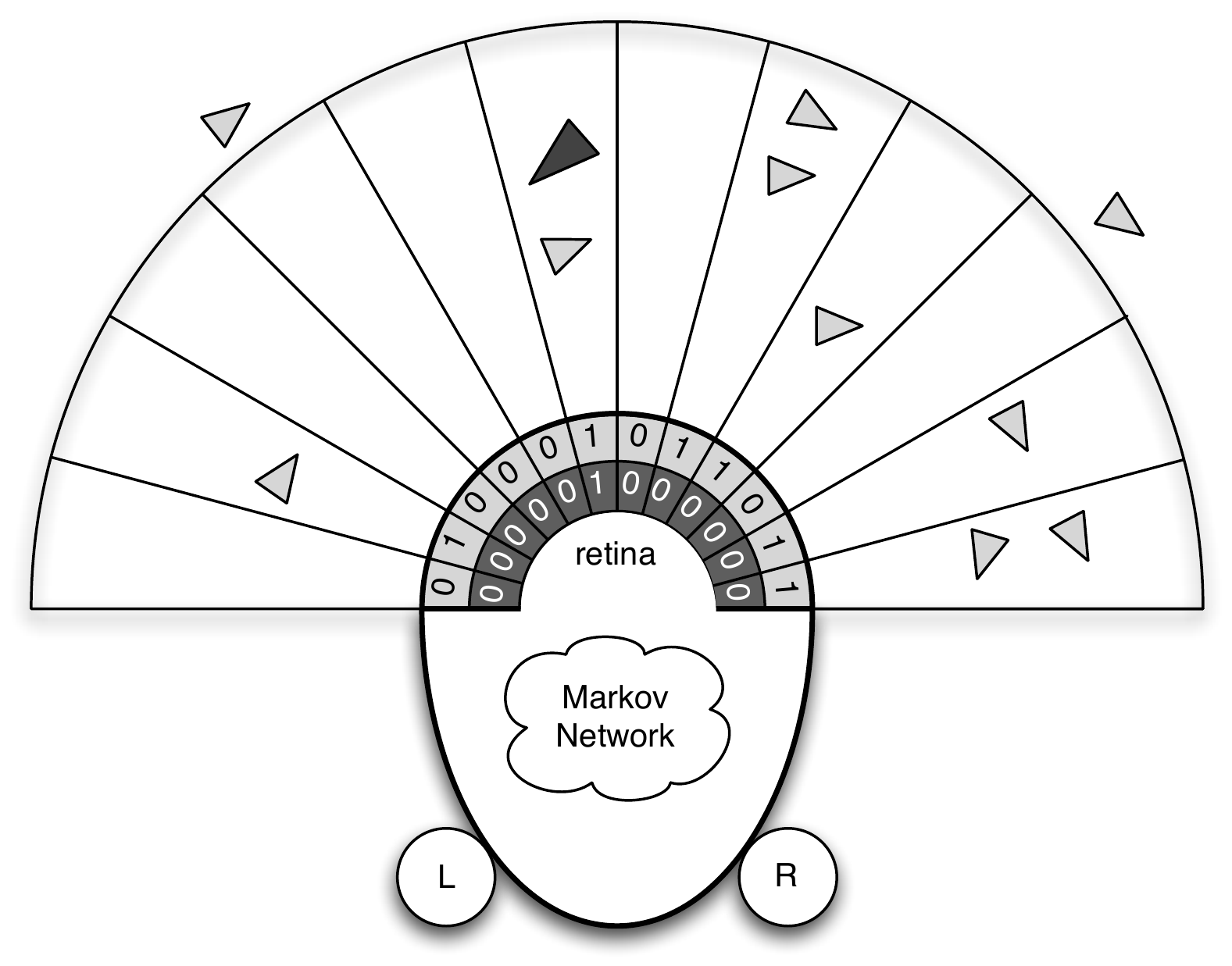}
\caption{An illustration of the predator and prey agents in the model. Light grey triangles are prey agents and the dark grey triangle is a predator agent. The predator and prey agents have a 180$^{\circ}$ limited-distance retina (100 virtual meters for the prey agents; 200 virtual meters for the predator agent) to observe their surroundings and detect the presence of the predator and prey agents. Each agent has its own Markov Network, which decides where to move next based off of a combination of sensory input and memory. The left and right actuators (labeled ``L" and ``R") enable the agents to move forward, left, and right in discrete steps.\label{fig:agent-illustration}}
\end{figure}

\subsection{Predator and prey agents}

Figure~\ref{fig:agent-illustration} depicts the sensory-motor architecture of predator and prey agents in this system. The retina sensors of both predator and prey agents are logically organized into ``layers," where a layer includes 12 sensors, with each sensor having a field of view of 15$^{\circ}$ and a range of 100 virtual meters (200 virtual meters for predators).  Moreover, each layer is attuned to sensing a specific type of agent.  Specifically, the predator agents have a single-layer retina that is only capable of sensing prey.  In contrast, the prey agents have a dual-layer retina, where one layer is able to sense conspecifics, and the other senses the predator.  (We note that there is only a single predator active during each simulation, hence the lack of a predator-sensing retinal layer for the predator agent.)


Regardless of the number of agents present in a single retina slice, the agents only know the agent type(s) that reside within that slice, but not how many, representing the wide, relatively coarse-grain visual systems typical in swarming birds such as Starlings~\cite{Martin1986}. For example in Figure~\ref{fig:agent-illustration}, the furthest-right retina slice has two prey in it (light grey triangles), so the prey sensor for that slice activates. Similarly, the sixth retina slice from the left has both a predator (dark grey triangle) and a prey (light grey triangle) agent in it, so both the predator and prey sensors activate and inform the MN that one or more predators \emph{and} one or more prey are currently in that slice. Lastly, since the prey near the 4th retina slice from the left is just outside the range of the retina slice, the prey sensor for that slice does not activate. We note that although the agent's sensors do not report the number of agents present in a single retina slice, this constraint does not preclude the agent's MN from evolving and making use of a counting mechanism which reports the number of agents present in a set of retina slices. Once provided with its sensory information, the prey agent chooses one of four discrete actions: (1) stay still; (2) move forward 1 unit; (3) turn left 8$^{\circ}$ while moving forward 1 unit; or (4) turn right 8$^{\circ}$ while moving forward 1 unit.

\begin{figure}
\centerline{\includegraphics[width=\HalfPage]{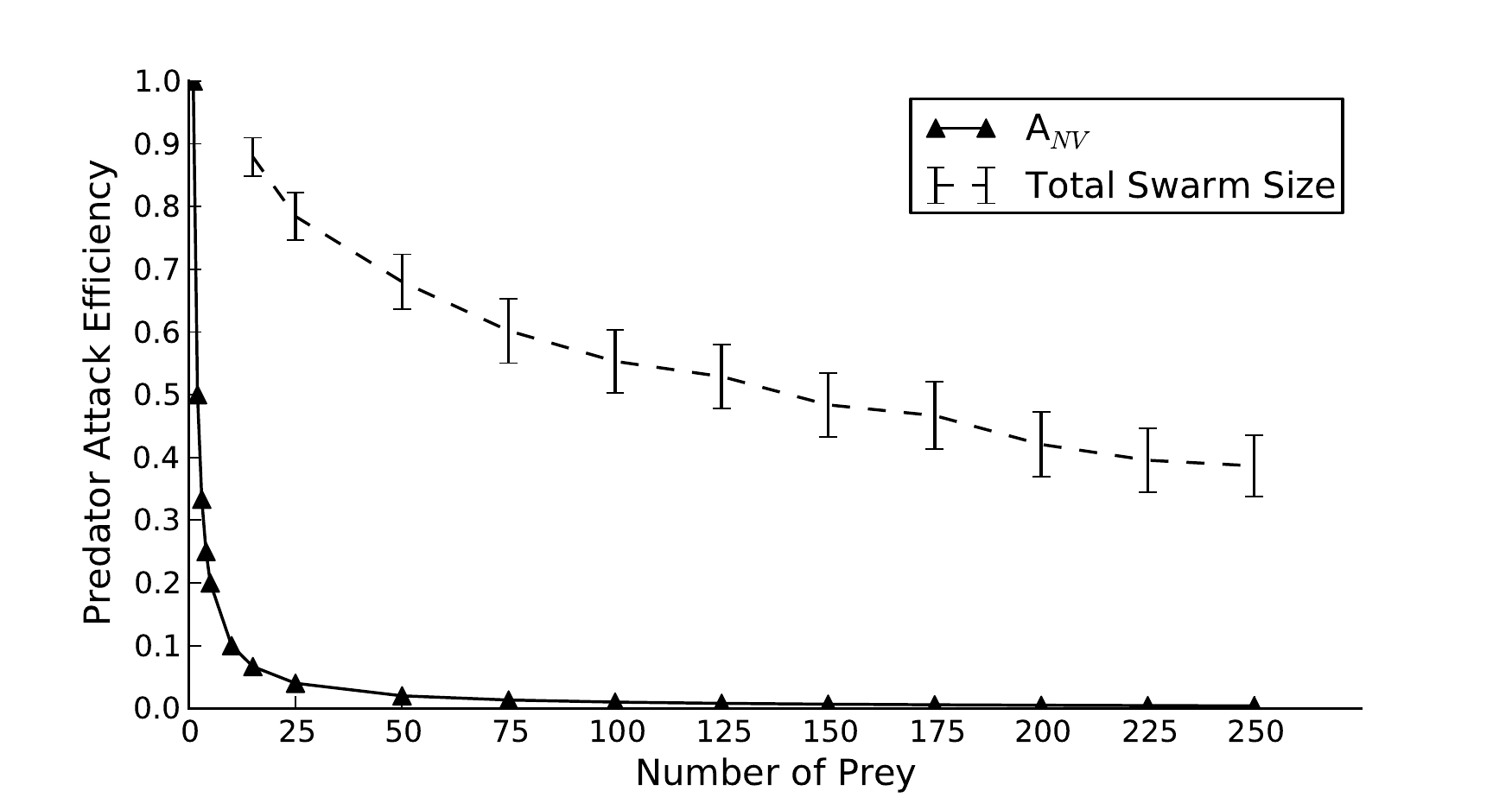}}
\caption{Relation of predator attack efficiency (\# successful attacks / total \# attacks) to number of prey. The solid line with triangles indicates predator attack efficiency as a function of the number of prey within the visual field of the predator ($A_{\rm{NV}}$). Similarly, the dashed line with error bars shows the actual predator attack efficiency given the predator attacks a group of swarming prey of a given size, using the $A_{\rm{NV}}$ curve to determine the per-attack predator attack success rate. Error bars indicate two standard errors over 100 replicate experiments.\label{fig:predator-attack-efficiency}}
\end{figure}

Likewise, the predator agent detects nearby prey agents using a limited-distance (200 virtual meters), pixelated retina covering its frontal 180$^{\circ}$ that functions just like the prey agent's retina. Similar to the prey agents, predator agents make decisions about where to move next, but the predator agents move 3x faster than the prey agents and turn correspondingly slower (6$^{\circ}$ per simulation time step) due to their higher speed.

\subsection{Simulation environment}

We use a simulation environment to evaluate the relative performance of the predator and prey agents. At the beginning of every simulation, we place a single predator agent and 50 prey agents at random locations inside a closed $512\times512$ unit two-dimensional simulation environment. Each of the 50 prey agents are controlled by clonal MNs of the particular prey MN being evaluated. We evaluate the swarm with clonal MNs to eliminate any possible effects of selection on the individual level, e.g., the ``selfish herd" effect~\cite{Olson2013,Wood2007}.


During each simulation time step, we provide all agents their sensory input, update their MN, then allow the MN to make a decision about where to move next. When the predator agent moves within 5 virtual meters of a prey agent it can see, it automatically makes an attack attempt on that prey agent. If the attack attempt is successful, the target prey agent is removed from the simulation and marked as consumed. Predator agents are limited to one attack attempt every 10 simulation time steps, which is called the \emph{handling time}. The handling time represents the time it takes to consume and digest a prey after successful prey capture, or the time it takes to refocus on another prey in the case of an unsuccessful attack attempt. Shorter handling times have negligible effects on the outcome of the experiment, except for when there is no handling time at all (Figure S2).

To investigate predator confusion as an indirect selection pressure driving the evolution of swarming, we implement a perceptual constraint on the predator agent. When the predator confusion mechanism is active, the predator agent's chance of successfully capturing its target prey agent ($P_{\rm{capture}}$) is diminished when any prey agents near the target prey agent are visible anywhere in the predator's visual field. This perceptual constraint is similar to previous models of predator confusion based on observations from natural predator-prey systems~\cite{Ioannou2008,Jeschke2007,Jeschke2005}, where the predator's \emph{attack efficiency} (\# successful attacks / total \# attacks) is reduced when attacking swarms of higher density. $P_{\rm{capture}}$ is determined by the equation:

\begin{equation}
P_{\rm{capture}} = \frac{1}{A_{\rm{NV}}}
\end{equation}
where $A_{\rm{NV}}$ is the number of prey agents that are visible to the predator, i.e., anywhere in the predator agent's visual field, \emph{and} within 30 virtual meters of the target prey. By only counting prey near the target prey, this mechanism localizes the predator confusion effect to the predator's retina, and enables us to experimentally control the strength of the predator confusion effect. Although our predator confusion model is based on the predator's retina, it is functionally equivalent to previous models that are based on the total swarm size (Figure~\ref{fig:predator-attack-efficiency}, dashed line), e.g., in~\cite{Ioannou2008,Jeschke2007,Jeschke2005,Tosh2006}. As shown in Figure~\ref{fig:predator-attack-efficiency} (solid line with triangles), the predator has a 50\% chance of capturing a prey with one visible prey near the target prey ($A_{\rm{NV}} = 2$), a 33\% chance of capturing a prey with two visible prey near the target prey ($A_{\rm{NV}} = 3$), etc. As a consequence, prey are in principle able to exploit the combined effects of predator confusion and handling time by swarming.

\section{Results}

\subsection{Effects of Predator Confusion}

\begin{figure}
\centerline{\includegraphics[width=0.35\textwidth]{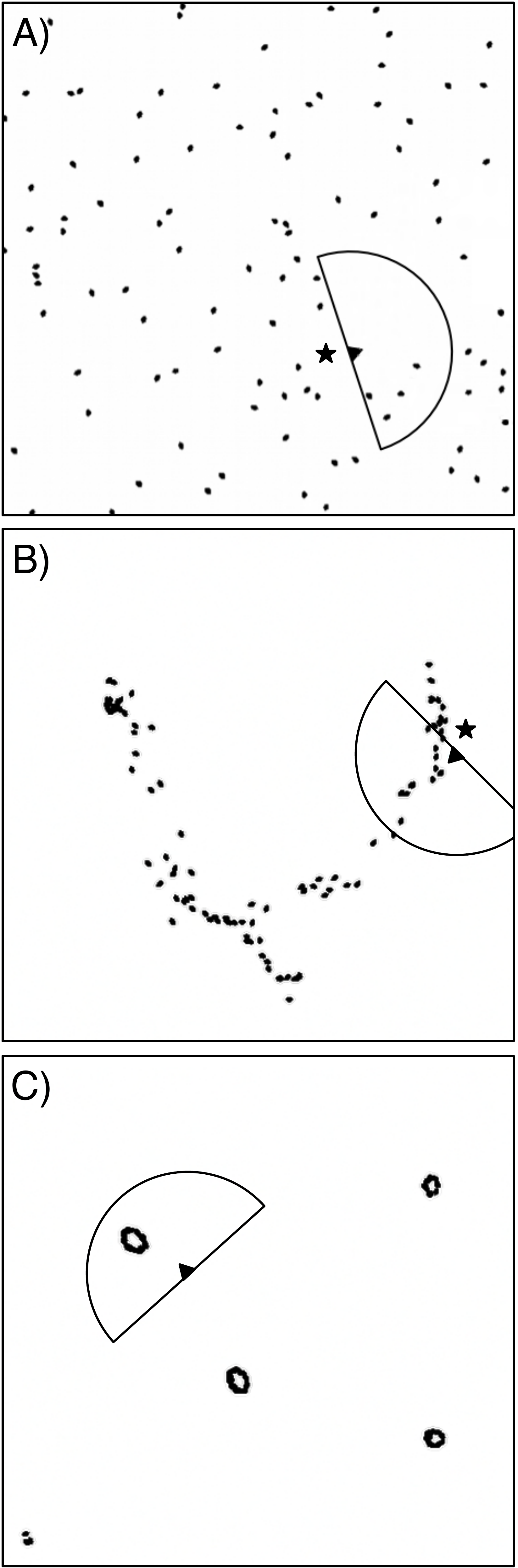}}
\caption{Screen captures of (A) dispersed prey in a swarm hunted by a predator without predator confusion, (B) prey forming a single elongated swarm under attack by a predator with predator confusion, and (C) prey forming multiple cohesive swarms to defend themselves from a predator with predator confusion after 1,200 generations of evolution. Black dots are prey, the triangle is the predator, the lines projecting from the predator represent the predator's frontal 180$^{\circ}$ visual field, and the star denotes where a prey was just captured.\label{fig:swarming-visualization}}
\end{figure}

Qualitatively, we observed significant differences in prey behavior over the course of evolution between swarms experiencing predators with and without predator confusion. Figure~\ref{fig:swarming-visualization}A illustrates that prey hunted by a predator without the predator confusion mechanism dispersed as much as possible to escape the predator. No replicates containing a predator without predator confusion resulted in prey behavior that resembled a cohesive swarm. Conversely, when evolution occurred with predator confusion, prey exhibited cohesive swarm behavior in the majority of the replicates (70\% of our replicates). Figure~\ref{fig:swarming-visualization}B depicts one such swarm in which prey follow the conspecific directly in front of them, resulting in an elongated swarm. Similarly, Figure~\ref{fig:swarming-visualization}C shows another swarm where the prey circle around their nearest conspecific, resulting in multiple small, cohesive swarms with the prey constantly trying to circle around each other. Both of these swarms evolved as defensive behaviors to exploit the predator confusion effect.


Furthermore, predators exhibited divergent hunting behaviors when hunting prey with and without predator confusion. As seen in Figure~\ref{fig:swarming-visualization}A, predators that evolved in the absence of predator confusion, and hence had to contend with dispersed prey, simply tracked the nearest visible prey until it was captured, then immediately pursued the next nearest visible prey. On the other hand, predators that evolved in the presence of predator confusion, and hence were challenged with cohesive swarms, used a mechanism that causes them to attack prey on the outer edges of the swarm. This strategy is similar to a predatory behavior observed in many natural systems~\cite{Hirsch2011,Romey2008}, and effectively minimized the number of prey in the predator's retina and maximized its chance of capturing prey. Figure~\ref{fig:swarming-visualization}B demonstrates this behavior, where the predator just captured a prey on the top-right edge of the swarm (prey capture location denoted by a black star). Videos of the evolved swarms under predation are available in the supplementary information (SI videos 1-5).

To evaluate the evolved swarms quantitatively, we obtained the line of descent (LOD) for every replicate by tracing the ancestors of the most-fit prey MN in the final population until we reached the randomly-generated ancestral MN with which the starting population was seeded (see~\cite{Lenski2003} for an introduction to the concept of a LOD in the context of digital evolution). For each ancestor in the LOD, we characterized the swarm behavior with two common behavior measurements: \emph{swarm density} and \emph{swarm dispersion}~\cite{Huepe2008}. We measured the swarm density as the mean number of prey within 30 virtual meters of each other over a lifespan of 2,000 simulation time steps. The swarm's dispersion was computed by averaging the distance to the nearest prey for every living prey over a lifespan of 2,000 simulation time steps. Together, these metrics captured whether or not the prey were cohesively swarming.

\begin{figure}
\centerline{\includegraphics[width=\HalfPage]{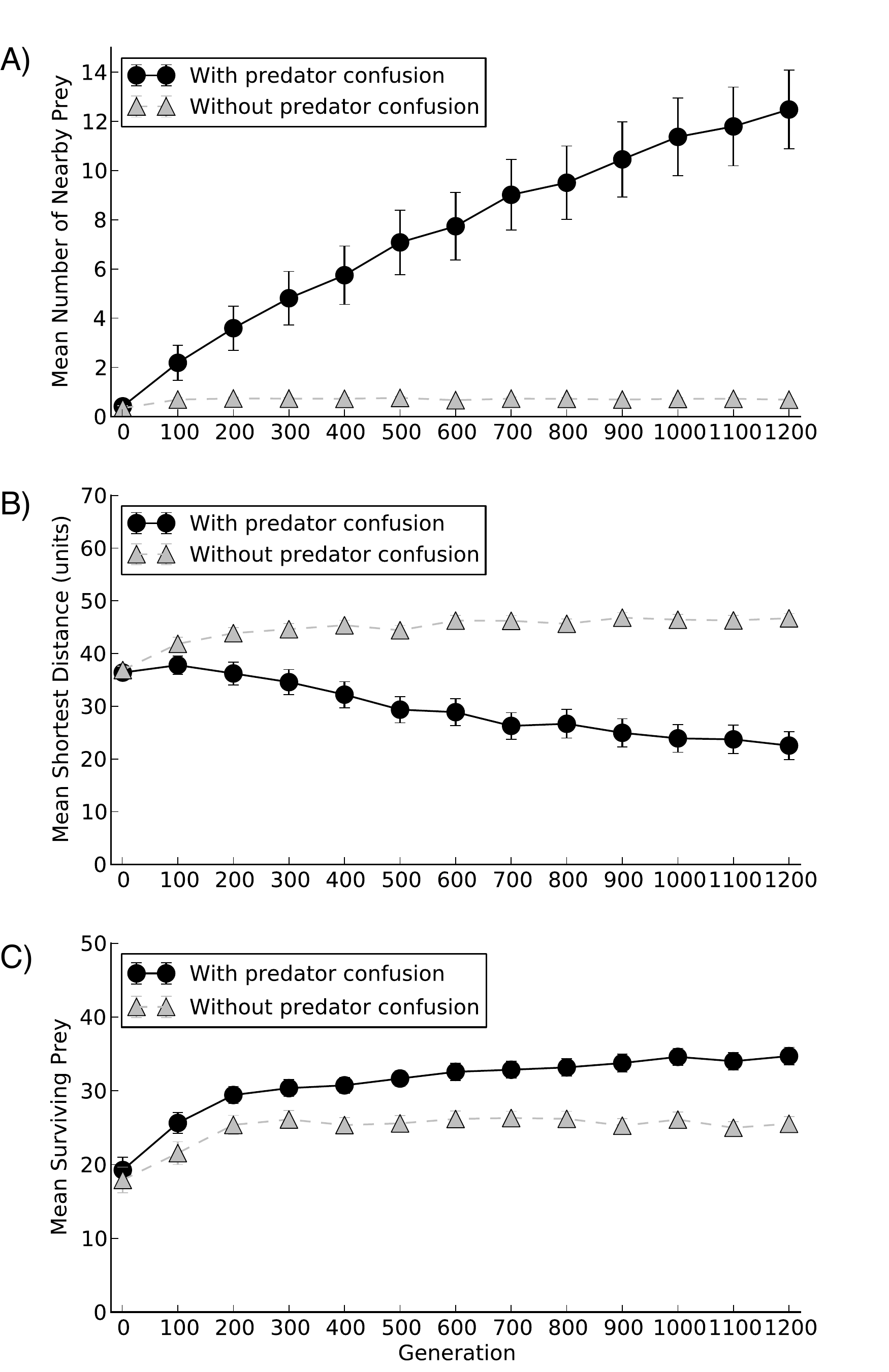}}
\caption{Mean swarm density (A), swarm dispersion (B), and survivorship (C) within the swarm over all replicates over evolutionary time. The swarm density was measured by the mean number of prey within 30 virtual meters of each other over a lifespan of 2,000 simulation time steps. Swarm dispersion was measured by the mean distance to the nearest prey for every living prey over a lifespan of 2,000 simulation time steps. Survivorship within the swarm was measured as the mean number of surviving prey (out of an initial total of 50) at the end of the simulation at a given generation. Prey hunted by a predator with predator confusion (black circles with a full line) evolved to maintain significantly higher swarm density and significantly less dispersed swarming behavior than prey in the swarms hunted by a predator without predator confusion (grey triangles with a dashed line). As a result, significantly more prey survived in the swarms hunted by a predator with predator confusion than the swarms hunted by a predator without predator confusion. Error bars indicate two standard errors across 180 replicate experiments.\label{fig:swarm-measurements}}
\end{figure}


Figure~\ref{fig:swarm-measurements}A demonstrates that the prey hunted by a predator with only handling time (i.e., without predator confusion) moved close to each other by chance but never coordinated their movement at any point in their evolutionary history (mean swarm density $\pm$ 1 standard error across 180 replicates: $0.69\pm0.02$). In contrast, when hunted by a predator with predator confusion, the prey coordinated their movement to remain close to each other and form a swarm (mean swarm density $12.48\pm0.8$ at generation 1,200). Likewise, Figure~\ref{fig:swarm-measurements}B shows that in the absence of predator confusion, prey evolved to maximize their dispersion (mean shortest distance $46.69\pm0.44$ at generation 1,200), whereas with predator confusion, prey evolved increasingly cohesive swarm behavior (mean shortest distance $22.54\pm1.32$ at generation 1,200). Taken together, these results confirm that predator confusion provided a sufficient selection pressure to evolve cohesive swarming behavior in this model, even though the swarming prey actually experience an increased attack rate from the predator due to this behavior (Figure S3 \& S4).


Figure~\ref{fig:swarm-measurements}C shows that as a result of these evolutionary trends, the cohesive swarms that evolved under predator confusion experienced significantly higher survivorship than swarms that evolved without predator confusion ($34.7\pm0.6$ and $25.54\pm0.49$ prey surviving the simulations, respectively). This increased survivorship confirms that swarming behavior confused the predator, leading to fewer successful prey captures. We found these results robust to a variety of experimental parameters, including weaker predator confusion effects (Figure S5 \& S6) and applying a minimum threshold to predator attack efficiency (Figure S7).


\subsection{Evolved Predator and Prey Behavior}

To deduce how swarms emerge in our model from individual-level behaviors, we next determined the functionality of the evolved predator and prey MNs. We accomplished this by first visualizing the MN connectivity to discern which slices of the retina and memory nodes of the MN were causally connected, then created a truth table from the MN mapping every possible input combination with its corresponding most-likely output from the MN. With this input-output mapping, we computed the minimal descriptive logic of the MN with Logic Friday, a hardware logic minimization program. We used the most-likely output for every input combination due to the stochastic nature of MNs, therefore the functionality we determined was the \emph{most-likely} behavior of the predator or prey.

In all of our experiments, the prey at generation 1,200 ignored the presence of predators and instead only reacted to the presence of conspecifics in their retina in order to follow the other prey in the swarm. This was particularly striking because it suggested that prey can evolve swarming behavior in response to predation without the ability to sense the predators hunting them, which was suggested in a previous study~\cite{Ioannou2012}. We observed that the prey evolved a wide variety of simple algorithms that exhibited a diversity of emergent swarming behaviors\footnote{Videos of the swarms are available as SI videos 2-5.}, ranging from moderately dispersed, elongated swarms similar to Starling murmurations (Figure~\ref{fig:swarming-visualization}B) to tighly-packed cohesive swarms reminiscent of fish bait balls (Figure~\ref{fig:swarming-visualization}C).

As for the predators, the evolved behavior we observed at generation 1,200 with predator confusion appeared to be rather complex: The predators avoided dense swarms and hunted prey outside, or on the edge, of the swarm. However, the algorithm underlying this behavior was relatively simple. The predators only watched the two center retina slices and constantly turned in one direction until a prey entered one of those slices. Once a prey became visible in one of the center retina slices, the predator moved forward and pursued that prey until it made a capture attempt. This process was repeated regardless of whether the predator successfully captured the prey. The simplicity of the predator algorithm and relative simplicity of the prey algorithms supports the findings of earlier digital swarm studies that complex swarm behaviors can be described by simple rules applied over a group of locally-interacting agents~\cite{Spector2005,Reynolds1987}.

\begin{figure}
\centerline{\includegraphics[width=\HalfPage]{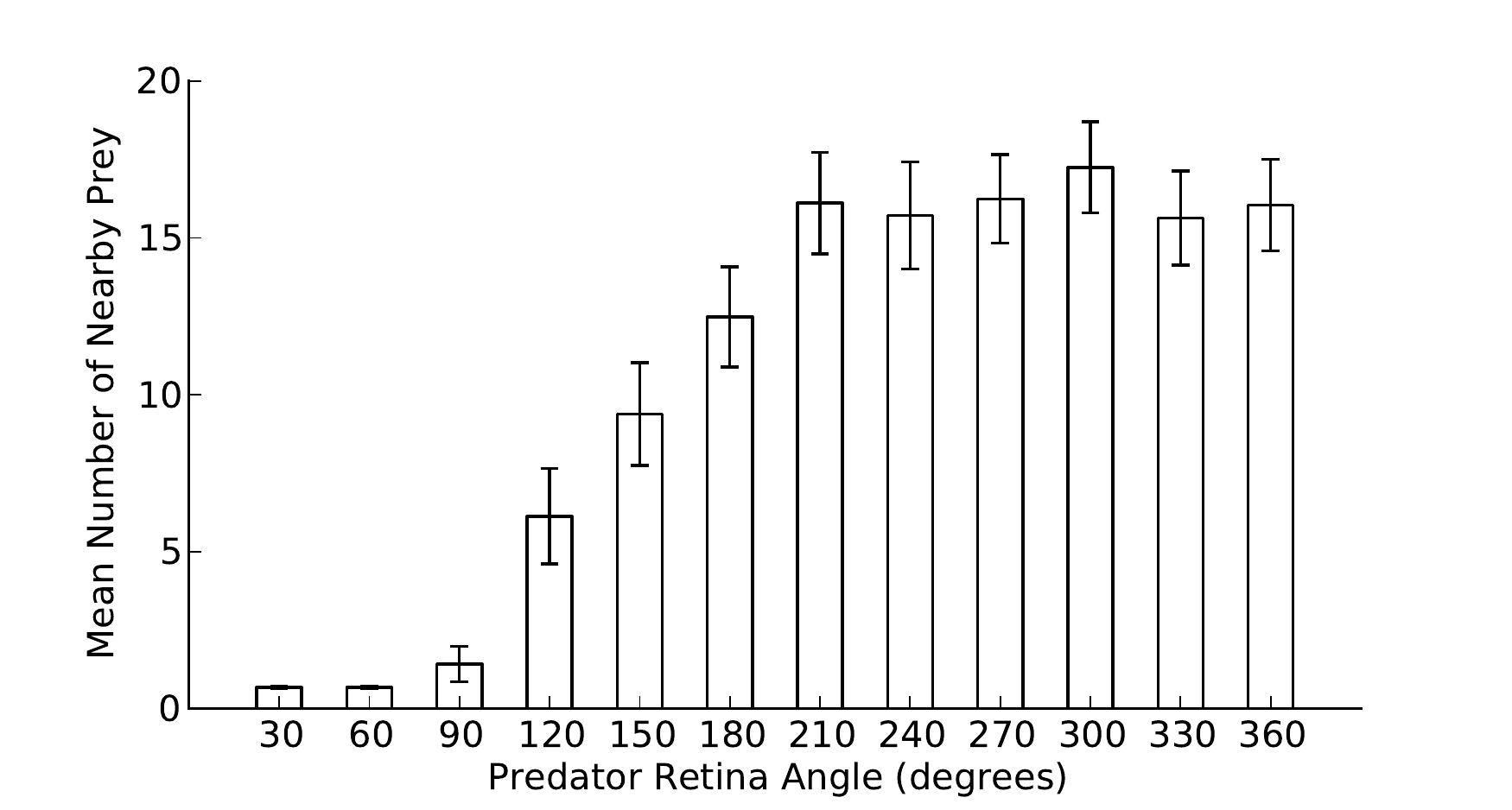}}
\caption{Mean swarm density at generation 1,200 as a function of predator view angle. Swarming to confuse the predator was an ineffective behavior if the predator's visual field covered only the frontal 60$^{\circ}$ or less, due to the predator's focused retina. As the predator's visual field was incrementally increased to cover the frontal 90$^{\circ}$ and beyond, predator confusion via swarming again became an effective anti-predator behavior, as evidenced by the swarms exhibiting significantly higher swarm density at generation 1,200. Error bars indicate two standard errors across 180 replicate experiments.\label{fig:swarm-density-pva}}
\end{figure}

\subsection{Effects of Predator Retina Angle}

We implemented predator confusion by imposing a perceptual constraint that reduces the probability of successfully capturing prey if one or more prey near the target prey are visible to the predator. This is meant to simulate the difficulty, arising from attentional or cognitive limitations, that a biological predator might have in choosing among multiple available prey at the moment of attack. To examine the effect of relaxing this constraint, we coevolved the predator and prey again and experimentally reduced the size of the predator's field of view. This procedure reduces the possibility that multiple prey can be detected at the moment of attack, thereby reducing the probability of confusion. For example, experimentally decreasing the predator's field of view from 180$^{\circ}$ to 60$^{\circ}$ decreases by two-thirds the area within which the presence of multiple prey can confuse the predator.

Figure~\ref{fig:swarm-density-pva} demonstrates that when the predator's retina only covered the frontal 60$^{\circ}$ or less, swarming to confuse the predator was no longer a viable adaptation (as indicated by a mean swarm density of $0.68\pm0.02$ at generation 1,200). In this case, the predator had such a narrow view angle that few swarming prey were visible during an attack, which minimizes the confusion effect and correspondingly increases its capture rate (Figure S8). As the predator's retina was incrementally modified to cover the frontal 120$^{\circ}$ and beyond, swarming again became an effective adaptation against the predator due to the confusion effect (indicated by a mean swarm density of $6.13\pm0.76$ at generation 1,200). This suggests that the predator confusion mechanism may not only provide a selective pressure for the prey to swarm, but it could also provide a selective pressure for the predator to narrow its view angle to become less easily confused.

\subsection{Effects on Functional Response}

\begin{figure}
\centerline{\includegraphics[width=\HalfPage]{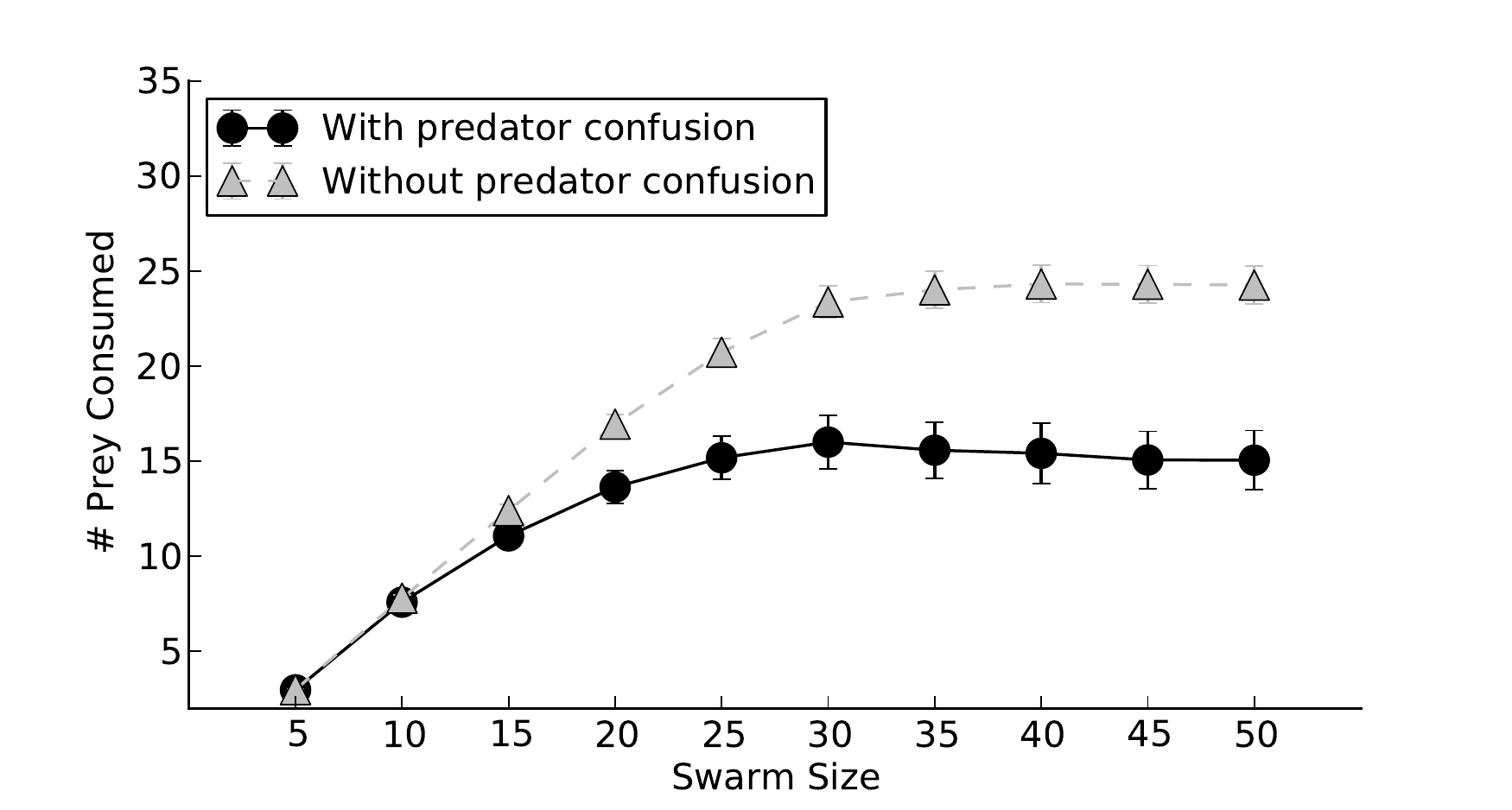}}
\caption{Functional response curves of cohesive swarms hunted by a predator with predator confusion (black circles with a full line) and dispersed swarms hunted by a predator without predator confusion (grey triangles with a dashed line). The evolved, cohesive swarms hunted by a predator with predator confusion result in a Type II functional response with a lowered plateau. Error bars indicate two standard errors across 180 replicate experiments.\label{fig:functional-response}}
\end{figure}

Predator confusion has been hypothesized to be not only a selective pressure favoring swarming, but also as a determinant of the \emph{functional response}~\cite{Jeschke2005}, i.e., the number of prey consumed by the predator as a function of prey density~\cite{Murdoch1973}. Figure~\ref{fig:functional-response} supports a key prediction of functional response theory: Both with and without predator confusion, the system displayed a Type II functional response (a saturating effect of prey density), but when predator confusion was present the functional response showed a lower plateau ($24.01\pm0.49$ prey consumed without predator confusion; $15.18\pm0.57$ with predator confusion). The fact that there was a Type II functional response even in the condition without predator confusion was the result of an additional constraint present in both conditions: The handling time that was imposed on the predator after prey capture before it can attack again. Additionally, when we varied the handling time in our experiments, we found that increasing the handling time also lowers the plateau of the Type II functional response (Figure S9).

\section{Discussion}

We demonstrated that swarming evolves as an emergent behavior in prey when a simple perceptual constraint---predator confusion---is imposed on the predator. Further, we found that measuring swarm density and swarm dispersion, proposed in~\cite{Huepe2008}, serves as an effective substitute for qualitatively assessing every swarm to determine if cohesive swarming behavior is present. A diverse collection of prey swarming behaviors evolved in our model, suggesting that predator confusion could allow for a wide range of swarming behaviors to evolve. Strikingly, most evolved prey strategies used algorithms that responded to other prey, but not to the attacking predators. This raises the interesting question of what selection pressures would favor the evolution of prey that detect and respond to the predators themselves.

In contrast to the diversity of evolutionary outcomes for prey, a common behavioral strategy emerged among the predators when evolved in the confusion condition. Namely, the evolved predators focused on attacking prey on the vulnerable edges of the swarms, which is a phenomenon commonly observed in nature~\cite{Hirsch2011,Romey2008}.

We also found that we could reduce the advantage of swarming by diminishing the predator's field of view, hence decreasing the level of confusion affecting the predator. This suggests that predator confusion could impose a selective pressure on the shape of the predator's retina: Once swarming has evolved in the prey, selection will favor predators that are no longer confused by swarms. Following the trend in Figure~\ref{fig:swarm-density-pva}, we would expect selection to favor predators with a narrower, more frontally focused retina, as observed in the visual systems of many natural predators~\cite{Tucker2000}.

Modeling functional response has been an important problem in ecology~\cite{Solomon1949}, and is critical for constructing accurate models that capture the dynamics of predator-prey interactions over ecological and evolutionary time~\cite{Jeschke2004}. We provided evidence that predator confusion has significant effects on functional response that are not captured in traditional models~\cite{Jeschke2005}. Most of these traditional models, including the original formulation of Holling~\cite{Holling1959}, capture the ecological interaction between predator and prey. Evolution is assumed to shape the behavioral strategies and constraints that influence predator-prey dynamics, but only recently have biologists begun to explicitly study the dynamics of predator-prey interactions over both ecological and evolutionary time~\cite{Hairston2005}. We have shown that a Type II functional response evolves even when it is not directly selected for, and the shape of the functional response can be attributed to specific constraints such as handling time and predator confusion.

\section{Conclusion}

We demonstrated that predator confusion provides a sufficient selective advantage for prey to evolve swarming behavior in a digital evolutionary model. This suggests that predator confusion likely contributed to the evolution of swarming behavior in animals which were hunted by predators that relied on visual systems to track their prey. Furthermore, in this work we (1) proposed a new method to directly test hypotheses about the evolution of swarming behavior, (2) provided an example of how to apply this method, and (3) demonstrated that by considering swarming behavior in the context of evolution, we are able to make discoveries about swarming behavior that were never previously considered. Of course, there are many other evolutionary pressures that have been hypothesized to lead to the evolution of swarming behavior~\cite{Krause2002}, such as the ``selfish herd" effect~\cite{Olson2013,Wood2007}, that remain to be explored in future work. Our results suggest that digital evolutionary systems can provide a powerful tool to tease apart these various hypothesized selective pressures underlying swarm behavior.

\section*{Acknowledgments}

This work was supported in part by the Paul G. Allen Family Foundation and the National Science Foundation BEACON Center for the Study of Evolution in Action under Cooperative Agreement DBI-0939454. We wish to acknowledge the support of the Michigan State University High Performance Computing Center and the Institute for Cyber Enabled Research (iCER).


\end{document}